\documentclass{article}

\usepackage{arxiv}
\usepackage[utf8]{inputenc} 
\usepackage[T1]{fontenc}    
\usepackage{booktabs}       
\usepackage{amsfonts}       
\usepackage{nicefrac}       
\usepackage{microtype}      
\usepackage{lipsum}		
\usepackage{amsmath,amssymb}
\usepackage{graphicx}
\usepackage{color}

\newcommand{\grl}{    {Geophys. Res. Lett.}}
\newcommand{\jgr}{    {J. Geophys. Res.}}
\newcommand{\ssr}{    {Space Sci. Rev.}}
\newcommand{\planss}{    {Plan. Sp. Sci.}}
\newcommand{\aap}{    { Astronomy and Astrophysics}}
\newcommand{\solphys}{ {Solar Physics}}
\newcommand{\apj}{ {Astrophys. J. }}
\newcommand{\apjl}{    {Astrophys. J. Lett.}}

\newcommand{\mnras}{ {Mon.Not.Royal.Soc. }}
\newcommand{\nat}{    {Nature}}

\newcommand{\prl}{    {Phys. Rev. Lett.}}
\newcommand{\apjs}{    {Astrophys. Journal. Suppl. Ser.}}

\newcommand{\blue}{\textcolor{black}}
\newcommand{\red}{\textcolor{black}}

\def\XXint#1#2#3{{\setbox0=\hbox{$#1{#2#3}{\int}$}
     \vcenter{\hbox{$#2#3$}}\kern-.5\wd0}}

\title{Solar wind current sheets: MVA inaccuracy and recommended single-spacecraft methodology}

\author{
	{Rachel Wang}\thanks{currently at Princeton University, New Jersey, USA} \\
	Space Sciences Laboratory, University of California at Berkeley, California, CA94720, USA\\
    \texttt{rw6831@princeton.edu} \\\\
    {Ivan Y. Vasko} \\
	Space Sciences Laboratory, University of California at Berkeley, California, CA94720, USA\\
	Space Research Institute of Russian Academy of Sciences, Moscow, 117997, Russia\\\\
	{Tai Phan} \\
	Space Sciences Laboratory, University of California at Berkeley, California, CA94720, USA\\\\
	\and
	{Forrest S. Mozer} \\
	Space Sciences Laboratory, University of California at Berkeley, California, CA94720, USA
}

\begin{document}
\maketitle
\begin{abstract}
We present the analysis of 2,033 current sheets (CS) observed aboard four Cluster spacecraft in a pristine solar wind. Four-spacecraft estimates of the CS normal and propagation velocity are compared with different single-spacecraft estimates. The Minimum Variance Analysis (MVA) of the magnetic field is shown to be highly inaccurate in estimating the normal. The MVA normal often differs by more than 60$^{\circ}$ from the normal obtained by multi-spacecraft timing method, likely due to ambient turbulent fluctuations. In contrast, the cross-product of magnetic fields at the CS boundaries delivers the normal with the uncertainty of less than 15$^{\circ}$ at the confidence level of 90\%. The CSs are essentially frozen into plasma flow, since their propagation velocity is consistent with local ion flow velocity within 20\% at the confidence level of 90\%. The single-spacecraft methodology based on the cross-product method and frozen-in assumption delivers the CS thickness and current density amplitude within 20\% of their actual values at the confidence level of 90\%. The CSs are kinetic-scale structures with half-thickness $\lambda$ from a few tenths to tens of local proton inertial length $\lambda_{p}$ and scale-dependent shear angle and current density amplitude, $\Delta \theta \propto (\lambda/\lambda_p)^{0.5}$ and $J_0\propto (\lambda/\lambda_{p})^{-0.5}$. The classification of the CSs in terms of tangential and rotational discontinuities remains a challenge, because even the four-spacecraft normal has too large uncertainties to reveal the actual normal magnetic field component. The presented results will be valuable for the analysis of solar wind CSs, when only single-spacecraft measurements are available.
\end{abstract}


\section{Introduction}


The early spacecraft measurements revealed the presence of directional discontinuities or, equivalently, current sheets in the solar wind at various radial distances from the Sun \blue{\cite{Burlaga69:multi_sc,Mariani73,burlaga77,tsurutani79,lepping86,Tsurutani96:jgr,soding01}}. The modern spacecraft measurements with higher temporal resolution demonstrated that solar wind current sheets are much more abundant than previously thought and have spatial thickness from a fraction to a few tens of proton inertial lengths \cite{Vasquez07,Perri12:prl,Podesta17:jgr,Artemyev18:apj,Vasko21:apj_rec,Vasko22:apj_origin,Lotekar22:apj}. It has been demonstrated that current sheets are substantially contributing into turbulence spectrum \cite{borovsky10,Podesta&Borovsky16:jgr} and correlated with enhanced electron and ion temperatures in the solar wind \cite{Borovsky&Denton11:apj,Osman12b,Wu13:apjl,Sioulas22:apj}. Even though a fraction of the strongest current sheets can potentially originate in solar corona \cite{Bruno01:pss,Borovsky08:jgr}, there is a strong evidence that the majority of solar wind current sheets are produced locally by turbulence cascade \cite{Vasquez07,Greco08,Greco09:apjl,Zhdankin12:apjl,Vasko22:apj_origin,Lotekar22:apj}. Similar coherent structures are indeed naturally formed in numerical simulations of plasma turbulence and substantially contribute into turbulence dissipation even though they occupy a relatively small volume \cite{Karimabadi13:phpl,Zhdankin13:apjl,Zhdanin14:apj,Wan12:prl,Wan16:phpl,Sisti21:aa,Jain21:apj}. Among the processes leading to turbulence dissipation is magnetic reconnection that has been observed in the solar wind \cite{Gosling07:grl_prevalence,Gosling&Szabo08,Mistry17,Phan06:nature,Phan09:grl,Phan10,Phan20:apjs,Eriksson22:apj} as well as turbulence simulations \cite{Servidio09:prl,servidio11,Cerri&Califano17,Franci17:apj,Papini19:apj}. The analysis of solar wind current sheets is therefore valuable not only for modelling solar wind turbulence, but also for understanding magnetic reconnection and plasma heating in turbulence in general.

Spacecraft measurements showed that the magnetic field across solar wind current sheets typically rotates through some \blue{shear} angle, but does not essentially vary in magnitude \blue{\cite{Burlaga69:multi_sc,Mariani73,burlaga77,tsurutani79,Vasquez07,Artemyev18:apj,Vasko21:apj_rec}}. The early studies were mostly based on single-spacecraft measurements, where current sheets were assumed locally planar one-dimensional structures frozen into the plasma flow (Taylor hypothesis), while current sheet normals were computed using Minimum Variance Analysis \cite{Mariani73,burlaga77,Neugebauer84,lepping86,Tsurutani96:jgr,soding01}. These studies were heavily focused on classifying solar wind current sheets in terms of tangential or rotational discontinuities based on the magnitude of the normal magnetic field component. Multi-spacecraft observations demonstrated however that Minimum Variance Analysis provides {\it highly inaccurate} estimates of current sheet normals in the solar wind \blue{and questioned thereby validity of these classifications} \cite{Horbury01,Knetter04}. Four-spacecraft Cluster measurements allowed estimating the normal by the timing method and showed that the normal magnetic field component is typically much smaller than local magnetic field magnitude and often within methodology uncertainties \cite{Knetter04}. Thus, solar wind current sheets can be tangential discontinuities and/or rotational discontinuities with a small normal \blue{component. The} relative occurrence of the different discontinuities in the solar wind is still not known \cite{Neugebauer06,Paschmann13,Artemyev19:grl,Artemyev19:jgr}. 

\blue{Knetter et al., 2004 \cite{Knetter04} showed} that the most accurate single-spacecraft estimate of the current sheet normal is delivered by the cross-product of magnetic fields at the current sheet boundaries. The recent statistical studies of \blue{various properties of} solar wind current sheets at 0.2 and 1 AU have \blue{been based on a single-spacecraft methodology relying} on the cross-product normals \cite{Vasquez07,Vasko21:apj_rec,Vasko22:apj_origin,Lotekar22:apj} and similar methodology has been widely used in the analysis of magnetic reconnection \cite{Gosling&Phan13,Phan10,Phan20:apjs,Mistry17,Eriksson22:apj}. This single-spacecraft methodology  needs to be thoroughly tested however, since even though Knetter et al., 2004 \cite{Knetter04} demonstrated the reliability of the cross-product normals, their analysis was limited to only about a hundred current sheets. \blue{Even more critical is that the selection procedure used by Knetter et al., 2004 \cite{Knetter04} was biased toward current sheets, whose thickness and shear angle were respectively larger than about 1,000 km and 30$^{\circ}$. The occurrence rate of these current sheets turned out to be less than one per hour. The recent studies showed however that solar wind current sheets at 1 AU are actually at least one order of magnitude more abundant and the most of current sheets have much smaller thickness and shear angle \cite{Vasquez07,Podesta17:jgr,Vasko21:apj_rec,Vasko22:apj_origin}.}

In this paper we present the analysis of more than two thousand current sheets observed aboard four Cluster spacecraft in a pristine solar wind. We substantially expand the previous multi-spacecraft studies by collecting a factor of ten more current sheets. \blue{Importantly, the collected dataset is not merely larger, but, in contrast to previous studies, also representative of solar wind current sheets typically present at 1 AU. Using this extensive and representative dataset we quantify} the accuracy of the estimates of current sheet properties by the single-spacecraft methodology based on the cross-product normal and frozen-in assumption \blue{(Taylor hypothesis)}. Using the collected extensive dataset we show that Minimum Variance Analysis (MVA) is indeed highly inaccurate in estimating current sheet normals independent of the theoretical indicator of MVA accuracy, the ratio between intermediate and minimum eigenvalues. We demonstrate that the single-spacecraft methodology used in the recent statistical studies delivers relatively accurate estimates of the current sheet properties including thickness and current density magnitude. The latter result is of value for modern studies of the solar wind near the Sun and beyond 1 AU, where only single-spacecraft measurements are available. Based on estimates of current sheet properties we discuss the nature and origin of current sheets in the solar wind.

\section{Data and methodology \label{sec:2}}

\section{Data and methodology \label{sec:2}}

\begin{figure}
\includegraphics[width=\textwidth]{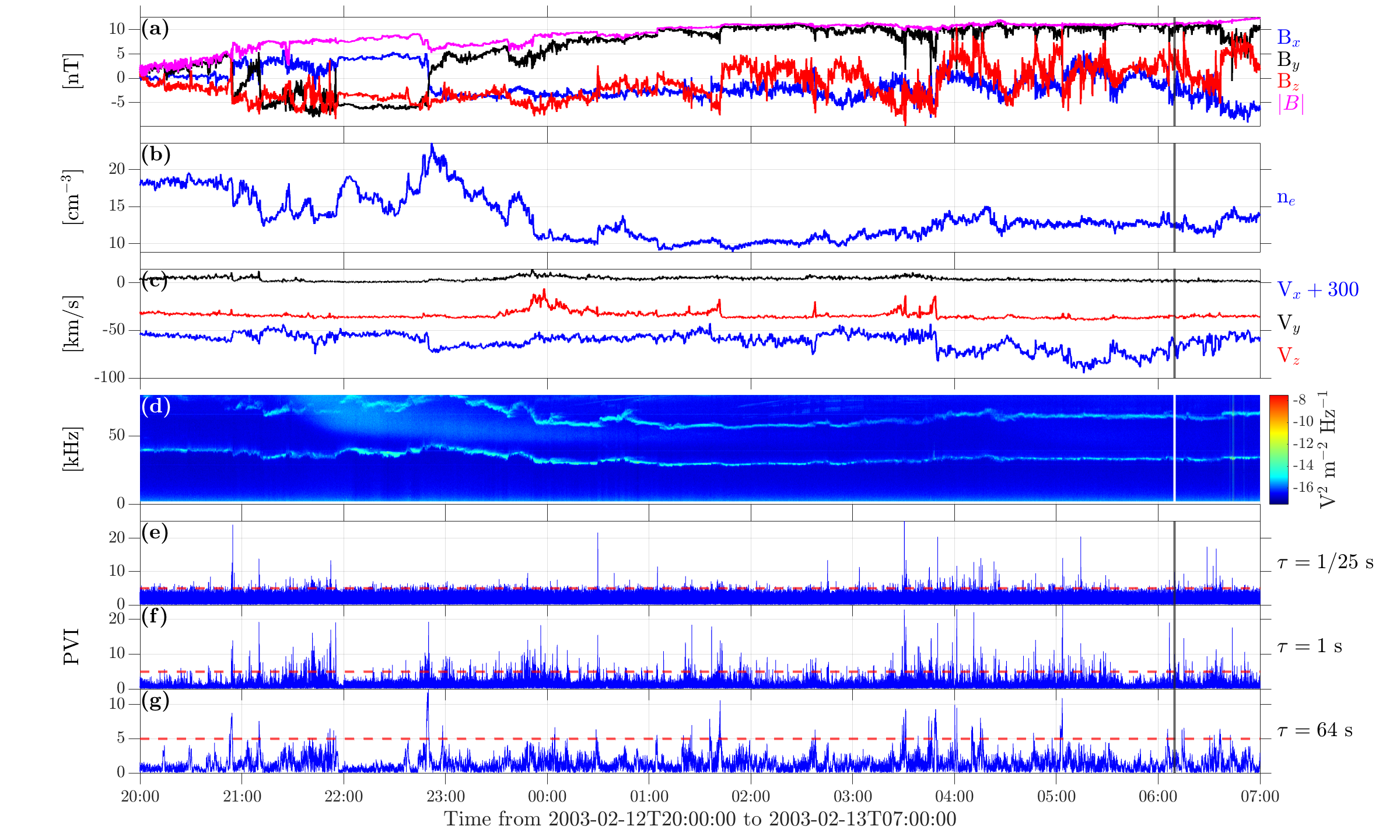}
\caption{Overview of Cluster observations in a pristine solar wind interval extending from February 12, 2003 to the next day. Panels (a)--(c) present C4 measurements of the magnetic field measured at \blue{22} Samples/s resolution, \blue{electron} density and flow velocity at \blue{8s} cadence, and electric field spectra at about 0.1s cadence; all vector quantities are in the Geocentric Solar Ecliptic (GSE) coordinates. Panels (e)--(g) demonstrate PVI indexes computed at time increments \blue{$\tau=1/22$}, 1 and 64s. Current sheets were collected by inspecting magnetic field fluctuations with PVI indexes above the threshold of ${\rm PVI}_{\tau}=5$ shown by red dashed lines. The vertical line around 06:10 UT on February 13 indicates the occurrence time of the current sheet shown in Figure \ref{fig2}.}
\label{fig1}
\end{figure}

We used the following data provided by Cluster spacecraft \cite{Escoubet01:angeo,Escoubet21:jgr}: magnetic field measurements at blue{22} S/s (Samples per second) resolution provided by Flux Gate Magnetometer \cite{Balogh01:angeo}, \blue{ion flow velocity at 4s cadence provided by CIS/HIA (Cluster Ion Spectrometry/Hot Ion Analyzer)} instrument \cite{Reme01:angeo}, electric field spectra from about 2 to 80 kHz at 0.1s cadence and electron density estimates at 2s cadence provided by the Whisper instrument \cite{Decreau01:angeo}. \blue{Note that we used proton flow velocity at 8s cadence provided by CIS/CODIF (Composition and Distribution Function) instrument \cite{Reme01:angeo} for the interval presented in Figure \ref{fig1}, because CIS/HIA moments were not available.} We visually inspected proton and electron pitch-angle distributions provided at a few second cadence respectively by CIS/CODIF \cite{Reme01:angeo} and Plasma Electron and Current Experiment \cite{Johnstone97:peace}. We also used proton temperature estimates provided by Wind spacecraft \cite{Wilson21}. Note that plasma density estimates provided by the Whisper instrument \blue{aboard Cluster} were consistent within 25\% with plasma density estimates provided by Wind spacecraft. Below the Geocentric Solar Ecliptic (GSE) coordinate system is notated $xyz$, four Cluster spacecraft are notated C1, C2, C3 and C4.

We selected current sheets from \red{221} pristine solar wind intervals from \red{35} days in 2003--2004 (Table \ref{tab:list}). The list of the intervals can be found in Supporting Materials (SM). A solar wind interval was considered pristine that is not affected by the Earth's bow shock if the electric field spectrum exhibited only electric field fluctuations at local plasma frequency and possibly its harmonics. Visually inspecting electron and proton pitch-angle distributions from a few tens of eV to about 30 keV, we assured the absence of electron and ion beams typical of foreshock. Table \ref{tab:list} presents typical values of background plasma parameters for each day along with maximum and minimum spatial separations between Cluster spacecraft. Note that Cluster spacecraft were separated by either less than 300 km or more than about 3,000 km.

The selection procedure of current sheets was based on a single-spacecraft Partial Variance of Increments (PVI) method, which allows identifying coherent structures by highlighting sharp gradients in the magnetic field compared to the background \cite{Greco08,Greco18}. The selection procedure was essentially equivalent to those used previously \cite{Podesta17:jgr,Vasko21:apj_rec,Vasko22:apj_origin}, but PVI indexes were computed for various time increments $\tau$, rather than only the increment dictated by data resolution. We recall that ${\rm PVI}_{\tau}(t)=\left(\sum_{\alpha} \Delta B^2_{\alpha}(t,\tau)/\sigma_{\alpha}^2\right)^{1/2}$, where $\Delta B_{\alpha}(t,\tau)=B_{\alpha}(t+\tau)-B_{\alpha}(t)$ are increments of three magnetic field components ($\alpha=x,y,z$) and $\sigma_{\alpha}$ is a standard deviation of $\Delta B_{\alpha}(t,\tau)$ computed over 2h intervals, a few outer correlation scales of solar wind turbulence \cite{Matthaeus05:prl}. We used magnetic field measurements aboard C4 to compute PVI indexes at time increments $\tau=$\blue{1/22, 4/22, 16/22}, 1, 4, 16, 64s, which allows identifying coherent structures of different scales. Each cluster of points satisfying ${\rm PVI}_{\tau}>5$ was encompassed by a 3 second to 3 minute interval depending on the time increment $\tau$ and the maximum variance component of the magnetic field computed by MVA for that interval was visually inspected \cite{Sonnerup&Scheible98}. Note that merging of overlapping intervals and exclusion of those nested into others have substantially reduced the required visual inspection. For all intervals with the maximum variance component reversing sign we manually adjusted current sheet boundaries, that is the regions to the left and right of the magnetic field reversal. Each current sheet identified aboard C4 was also observed aboard other Cluster spacecraft, though a few tens of current sheets had to be excluded because the magnetic field profile observed aboard one of the other spacecraft was not correlated well enough with the magnetic field profile observed aboard C4. The final dataset includes \red{2,033} current sheets observed aboard four Cluster spacecraft (Table \ref{tab:list}).

Figure \ref{fig1} presents an overview of the longest interval from our dataset, encompassing February 12 and 13, 2003. Panels (a)--(c) present the magnetic field, plasma density and flow velocity observed aboard C4. The magnetic field magnitude was 2--10 nT, the plasma density and proton flow velocity were respectively around 10--20 cm$^{-3}$ and 350 km/s through the entire interval. The electric field spectrum in panel (d) indicates that Cluster was in a pristine solar wind, since electric field fluctuations were only observed at local electron plasma frequency and its second harmonic. PVI indexes computed at several time increments, \blue{$\tau=1/22$}, 1 and 64s, are presented in panels (e)--(g). In accordance with higher intermittency at smaller scales \cite{Bruno19:intermit}, we observe higher occurrence of ${\rm PVI}_\tau>5$ at smaller time increments. This indicates higher occurrence of smaller scale coherent structures in the solar wind. Also in accordance with previous observations \cite{Greco16:apjl}, small-scale coherent structures are typically around large-scale coherent structures. Because of that almost all current sheets are already identified by ${\rm PVI}_\tau>5$ at several smallest time increments.

\begin{figure}
\noindent\includegraphics[width=\textwidth]{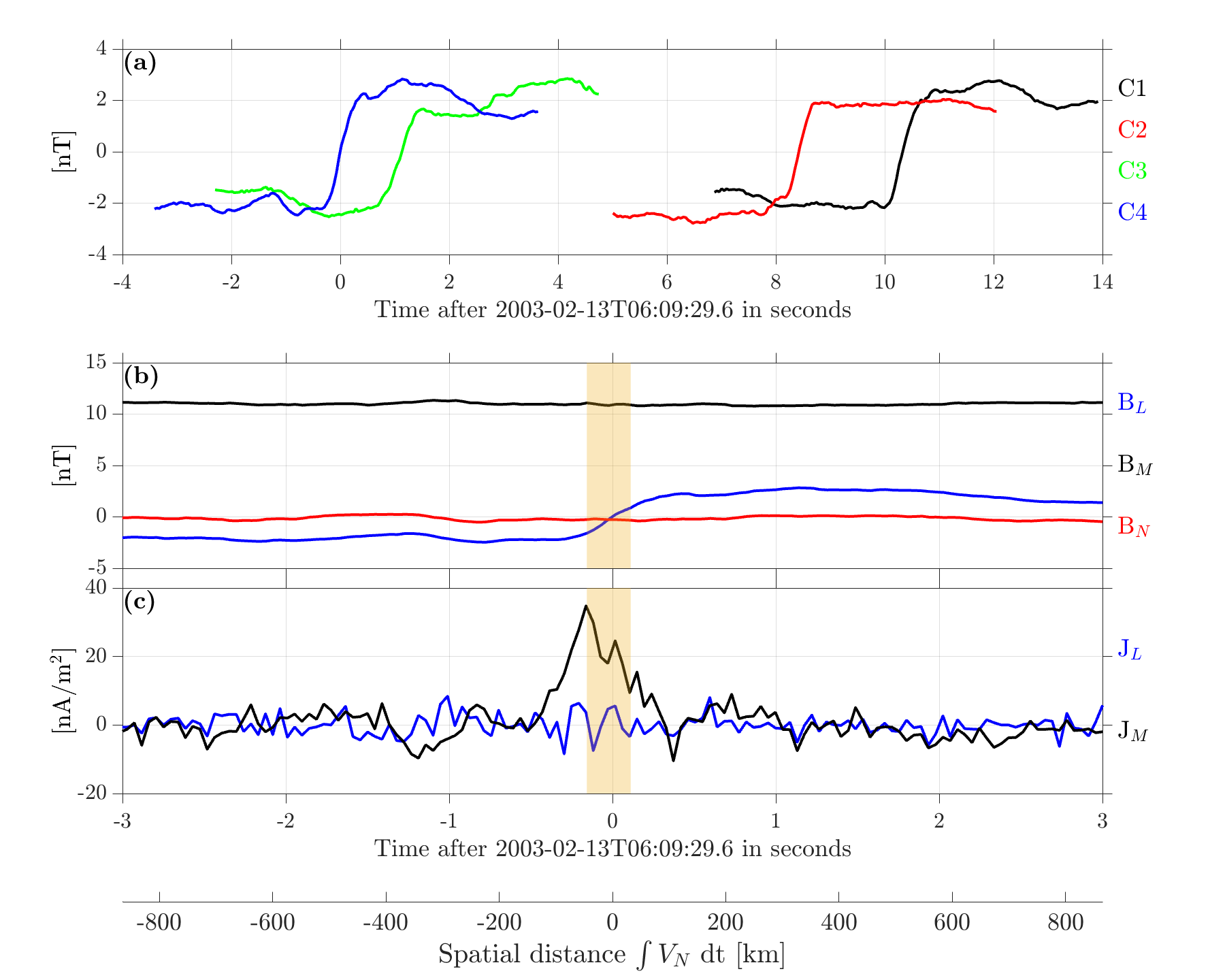}
\caption{The current sheet observed on February 13, 2003 around 06:10 UT (Section \ref{sec:2}): (a) the magnetic field component $B_{L}$ observed aboard C1--C4 at 25 Samples/s resolution, where ${\bf L}$ is the optimal maximum variance direction of the magnetic field; (b) three magnetic field components observed aboard C4 in the local coordinate system; $B_{L}$ is the maximum variance component, $B_{M}$ and $B_{N}$ are guide field and normal components; (c) current density components computed using the current sheet propagation velocity $V_{N}$ revealed by the timing method, $J_{M}=-\left(\mu_0 V_{N}\right)^{-1}dB_{L}/dt$ and $J_{L}=\left(\mu_0 V_{N}\right)^{-1}dB_{M}/dt$, \blue{where the time step is dictated by data resolution, $dt=1/22$ s. The CS central region is highlighted in panels (b) and (c) and corresponds to $|B_{L}-\langle B_{L}\rangle| \leq 0.25\Delta B_{L}$, where $\Delta B_{L}$ is the variation of $B_{L}$ between the boundaries and $\langle B_{L}\rangle$ is the mean of $B_{L}$ values at the CS boundaries.} The bottom axis converts temporal scale of the current sheet into spatial scale by multiplying with the revealed propagation velocity.}
\label{fig2}
\end{figure}


Figure \ref{fig2} presents a current sheet (CS) observed around 06:10 UT on February 13, 2003 to demonstrate our methodology. Panel (a) presents the magnetic field component $B_L$ observed aboard individual Cluster spacecraft. Note that ${\bf L}\approx (0.58,-0.16, 0.80)$ is the optimal maximum variance vector, that is a unit vector with the least angular deviation from four maximum variance vectors estimated by applying MVA to magnetic field measurements of individual Cluster spacecraft. The maximum angle between individual maximum variance vectors of $\alpha_{LL}\approx 2^{\circ}$ indicates a local CS planarity. Applying timing or triangulation technique \cite{Burlaga69:multi_sc,Horbury01,Knetter04,Wang22:IHs}, which uses time delays between the occurrence of the CS aboard individual Cluster spacecraft, we estimate the local CS normal and its propagation velocity along the normal in the spacecraft frame. The revealed timing normal ${\bf N}_{\rm T}=(-0.77, -0.40, 0.49)$ is almost perpendicular to the optimal maximum variance vector ${\bf L}$ with corresponding angle of $\alpha_{NL}\approx 88^{\circ}$, which is another indication of the local CS planarity. The angle $\alpha_{\rm max}={\rm max}(\alpha_{LL},|90^{\circ}-\alpha_{NL}|)\approx 2^{\circ}$ can be considered a quantitative indicator of CS planarity on the scale of Cluster spacecraft separation. The revealed CS propagation velocity $V_{N}\approx 284$ km/s is quite consistent with the normal component of the ion flow velocity $V_{iN}\equiv {\bf V}_{i}\cdot {\bf N}_{\rm T}\approx 297$ km/s, which reflects the validity of the Taylor hypothesis.

Applying MVA to magnetic field measurements of individual Cluster spacecraft we computed four minimum variance vectors. We only use minimum variance vector ${\bf N}_{\rm MVA}=(0.40,-0.79,-0.47)$ revealed aboard C4, but the use of the other normal vectors results in essentially identical results (not shown). The eigenvalues corresponding to maximum, intermediate and minimum variance vectors were  $(\lambda_{\rm max},\lambda_{\rm int},\lambda_{\rm min})\approx (4,0.04,0.01)$. The minimum variance vector ${\bf N}_{\rm MVA}$ is almost perpendicular to the timing normal ${\bf N}_{\rm T}$, the corresponding angle is about $110^{\circ}$. This analysis demonstrates that MVA is {\it highly inaccurate} in estimating the CS normal in spite of relatively high ratio between intermediate and minimum eigenvalues, $\lambda_{\rm int}/\lambda_{\rm min}\approx 4$. In contrast, the cross-product of averaged magnetic fields at the CS boundaries delivers vector ${\bf N}_{\rm X}=(-0.77,-0.37,0.52)$ that is consistent with the timing normal ${\bf N}_{\rm T}$ within about $2^{\circ}$. 

The magnetic field in panel (b) is presented in a local CS coordinate system ${\bf LMN}$, where ${\bf M}={\bf N}_{\rm T}\times {\bf L}$ is the guide field direction, while the maximum variance vector ${\bf L}$ was corrected, ${\bf L}\rightarrow {\bf L}-{\bf N}_{\rm T}\left({\bf L}\cdot {\bf N}_{\rm T}\right)$, to make it strictly perpendicular to the timing normal and normalized to unity. Note that, in principle, we could correct the timing normal instead of the maximum variance vector. The magnetic field magnitude was around 11 nT across the CS, while the normal component $B_{N}$ was only about 0.1 nT within the CS. The revealed normal component of 0.1 nT is however within the methodology uncertainty of $11\;{\rm nT}\cdot{\rm sin}\;\alpha_{\rm max}\approx 0.4$ nT resulting from the angular uncertainty $\alpha_{\rm max}\approx 2^{\circ}$ of the maximum variance vector and timing normal. This angular uncertainty results from the fact that we could, in principle, correct the timing normal to make it perpendicular to the optimal maximum variance vector and also use the maximum variance vectors observed aboard individual Cluster spacecraft rather than the optimal vector ${\bf L}$ (SM). The current sheet is therefore either a tangential discontinuity or rotational discontinuity with the normal component within the methodology uncertainty of 0.4 nT. Importantly, the use of minimum variance vector ${\bf N}_{\rm MVA}$ for the CS normal would overestimate the normal component by a factor of hundred, since MVA basically confuses the normal and guide field directions for this CS.

We use the propagation velocity of the CS to translate its temporal magnetic field profile into spatial one and also compute the current density presented in panel (c): $J_{M}=-\left(\mu_0 V_{N}\right)^{-1}dB_{L}/dt$ and $J_{L}=\left(\mu_0 V_{N}\right)^{-1}dB_{M}/dt$, where $\mu_0$ is the vacuum permeability \blue{and the time step is dictated by data resolution, $dt=1/22$ s}. Note that the curlometer technique would substantially underestimate the current density (e.g., Vasko et al., 2014 \cite{Vasko14:tiltedCS}), because the thickness of solar wind CSs is typically smaller than Cluster spacecraft separation \cite{Vasquez07,Vasko21:apj_rec,Vasko22:apj_origin}. The CS half-thickness is estimated as $\lambda=\Delta B_{L}/2\mu_0\langle J_M\rangle$, where $\Delta B_{L}$ is the variation of $B_{L}$ between the CS boundaries and $\langle J_M\rangle$ is the current density averaged over the CS central region. The latter is highlighted in the panels and corresponds to $|B_{L}-\langle B_{L}\rangle| \leq 0.25\Delta B_{L}$, where $\langle B_{L}\rangle$ is the mean of $B_{L}$ values at the CS boundaries \blue{and the factor of 0.25 has been chosen to have at least three data points in the central region for all the CSs in our dataset}. Since solar wind CSs are predominantly magnetic field rotations and the current density is dominated by the component parallel to local magnetic field \cite{Vasquez07,Vasko21:apj_rec,Vasko22:apj_origin}, we characterize each CS by the parallel current density averaged over the CS central region, $J_0\equiv \langle J_{||}\rangle$ 
and shear angle $\Delta\theta$ between magnetic fields at the CS boundaries. We found that the considered CS is a proton kinetic-scale structure with half-thickness $\lambda\approx 65$ km that is around $\lambda_{p}$ in units of local proton inertial length. The averaged parallel current density $J_0\approx 24$ nA/m$^{2}$ is around 0.2$J_{A}$, where $J_{A}=en_{i}V_{A}$ is local Alfv\'{e}n current density and $V_{A}$ is local Alfv\'{e}n speed. Both proton inertial length and Alfv\'{e}n current density are so-called Alfv\'{e}n units typically used in turbulence simulations \cite{Franci17:apj,Papini19:apj,Jain21:apj}.



We carried out similar analysis for all the collected 2,033 CSs. We obtained single-spacecraft estimates of the half-thickness and current density notated respectively $\lambda^{*}$ and $J_0^{*}$. In the single-spacecraft methodology that was applied to C4 measurements the maximum variance vector is still delivered by MVA, but the CS normal is determined by the cross-product method and the propagation velocity is assumed to coincide with the normal component of local ion flow velocity \cite{Vasquez07,Phan10,Vasko22:apj_origin}. Single-spacecraft estimates ($\lambda^{*}$ and $J_0^{*}$) will generally differ from four-spacecraft estimates ($\lambda$ and $J_0$) because the normal and propagation velocity are computed differently. Note that for 28 CSs the magnetic field profile was bifurcated and the current sheet thickness and current density were not computed, because the current density $\langle J_{M}\rangle$ averaged over the central region of a bifurcated CS is close to zero and the estimate $\Delta B_{L}/2\mu_0\langle J_{M}\rangle$ does not reflect the actual CS thickness \cite{Vasko21:apj_rec}. These CSs were only excluded from the analysis of the current sheet thickness and current density amplitude in Figure \ref{fig9}.  In the next section we quantify the accuracy of the different methods in estimating the actual CS normal and also the accuracy of the single-spacecraft methodology in estimating the CS properties.

\begin{figure}
\noindent\includegraphics[width=\textwidth]{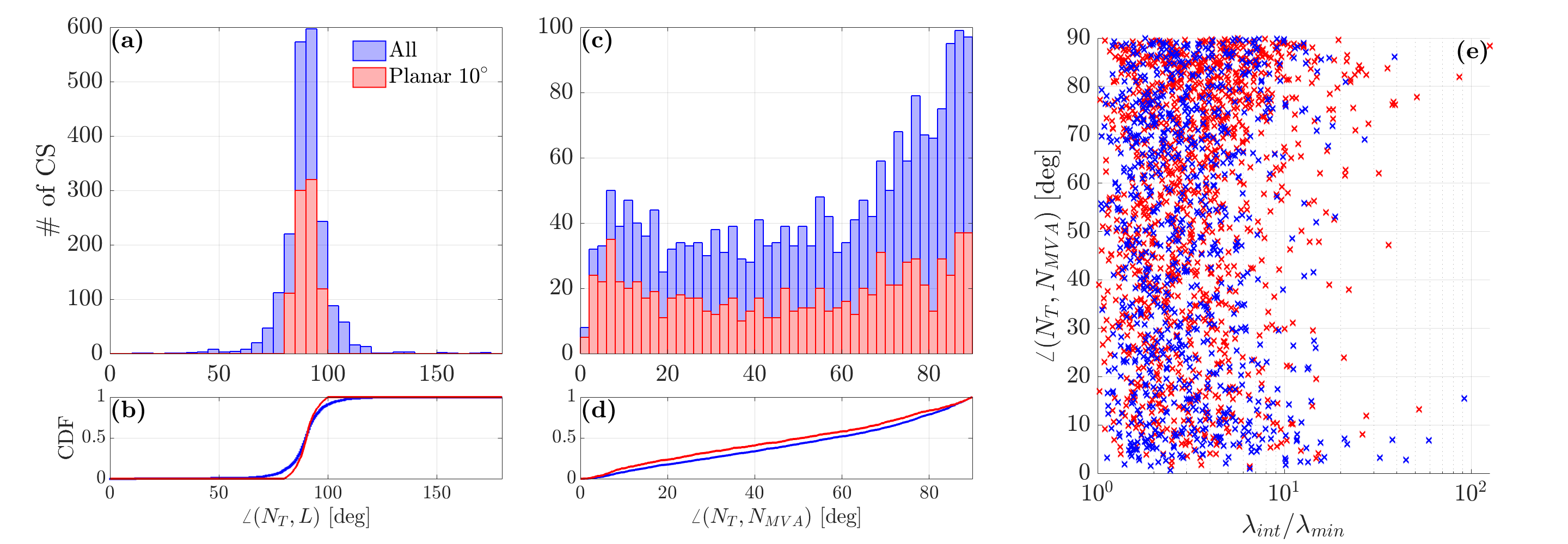}
\caption{The demonstration of planarity of the CSs and MVA inaccuracy in estimating the CS normal (Section \ref{sec:3}). Panels (a) and (b) present probability and cumulative distributions of the angle between optimal maximum variance vector ${\bf L}$ and timing normal ${\bf N}_{\rm T}$. Panels (c) and (d) show probability and cumulative distributions of the angle between timing normal ${\bf N}_{\rm T}$ and minimum variance vector ${\bf N}_{\rm MVA}$ computed for the magnetic field measured aboard C4. Panel (e) demonstrates a scatter plot of the angle between ${\bf N}_{\rm MVA}$ and ${\bf N}_{\rm T}$ versus the ratio $\lambda_{\rm int}/\lambda_{\rm min}$ between intermediate and minimum eigenvalues that is supposed to be a theoretical indicator of MVA accuracy. The distributions and data shown by different colors correspond to all CSs and CSs of sufficiently high planarity with $\alpha_{\rm max}\leq 10^{\circ}$.}
\label{fig3}
\end{figure}

\section{MVA and recommended single-spacecraft methodology \label{sec:3}}

Figure \ref{fig3} presents the analysis of MVA accuracy in estimating the actual CS normal. Importantly, the actual CS normal is not known exactly, since the timing normal has uncertainties associated at least with the degree of CS planarity on the scale of Cluster spacecraft separation. The analysis of the CSs of sufficiently high planarity allows however improving the alignment of the timing normal with the actual CS normal. Panels (a) and (b) present probability and corresponding cumulative distributions of the angle $\alpha_{NL}$ between the optimal maximum variance vector ${\bf L}$ and the timing normal ${\bf N}_{\rm T}$. Typically the two vectors are almost perpendicular as it should be for locally planar structures. We also demonstrate in panels (a) and (b) similar distributions for CSs of different degree of planarity quantified by $\alpha_{\rm max}={\rm max}\left(|90^{\circ}-\alpha_{NL}|,\alpha_{LL}\right)$, where we recall that $\alpha_{LL}$ is the maximum angle between maximum variance vectors observed aboard individual Cluster spacecraft. Not surprisingly the distribution of $\alpha_{NL}$ is more narrow around 90$^{\circ}$ for the CSs of higher planarity. Panels (c) and (d) present probability and cumulative distributions of the angle between the minimum variance vector ${\bf N}_{\rm MVA}$ observed aboard C4 and the timing normal ${\bf N}_{\rm T}$; the use of minimum variance vectors observed aboard other Cluster spacecraft would result in similar statistical distributions (not shown). The MVA turns out to be {\it highly inaccurate} in estimating the CS normal, since the minimum variance vector is typically highly oblique to the timing normal. The corresponding angle is often around 90$^{\circ}$ and larger than $60^{\circ}$ for about 50\% of the CSs. The MVA accuracy is not strongly dependent on the degree of local CS planarity, since probability and cumulative distributions for the CSs with $\alpha_{\rm max} \leq 10^{\circ}$ are essentially identical. Even more surprising panel (e) shows that the MVA accuracy is not correlated with the ratio $\lambda_{\rm int}/\lambda_{\rm min}$ between intermediate and minimum eigenvalues that is theoretically supposed to indicate MVA accuracy or at least be a requirement for accuracy \cite{Behannon80:jgr,Sonnerup&Scheible98}.

\begin{figure}
\noindent\includegraphics[width=0.75\textwidth]{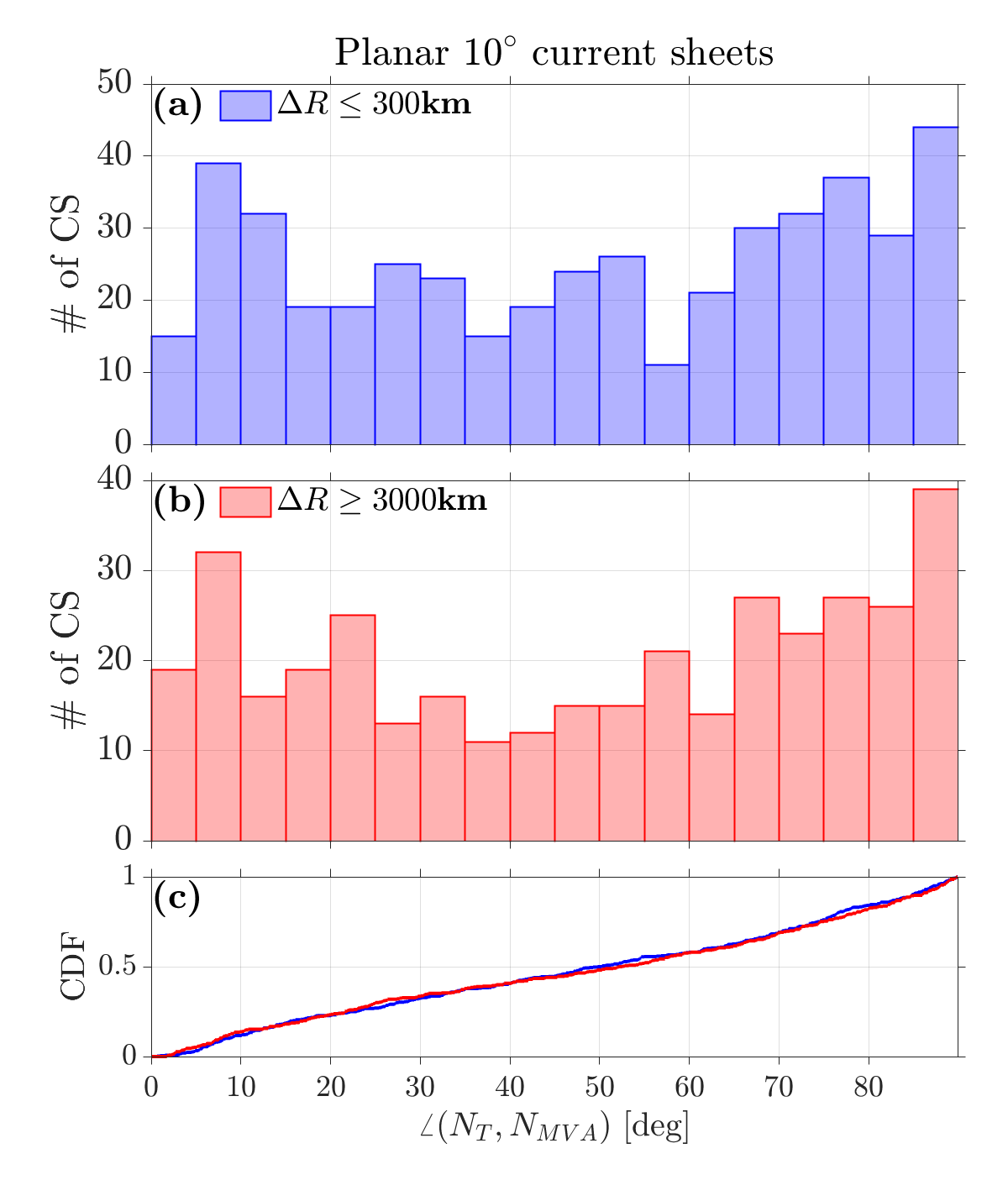}
\caption{The demonstration that MVA accuracy is independent of the maximum spatial separation between Cluster spacecraft: (a,b) probability distributions of the angle between the minimum variance vector ${\bf N}_{\rm MVA}$ and the timing normal ${\bf N}_{\rm T}$ for the CSs of high planarity, $\alpha_{\rm max}\leq 10^{\circ}$, but observed in the intervals, when the maximum spatial separation between Cluster spacecraft was either less than 300 km or larger than 3,000 km; (c) cumulative distribution functions corresponding to the probability distributions.}
\label{fig4}
\end{figure}

Figure \ref{fig4} presents the analysis of MVA accuracy dependence on the separation between Cluster spacecraft. Panels (a) and (b) show probability distributions of the angle between ${\bf N}_{\rm MVA}$ and ${\bf N}_{\rm T}$ for the CSs of high planarity, $\alpha_{\rm max}\leq 10^{\circ}$, but observed in the intervals, when the maximum spatial separation between Cluster spacecraft was either less than 300 km or larger than 3,000 km. Along with the corresponding cumulative distributions in panels (c) these probability distributions show that MVA inaccuracy is independent of the spatial separation between Cluster spacecraft. This statement is also corroborated by the analysis of all the CSs, rather than only those of  sufficiently high planarity (SM). We speculate on the cause of the revealed MVA inaccuracy in Section \ref{sec:4}.

\begin{figure}
\noindent\includegraphics[width=\textwidth]{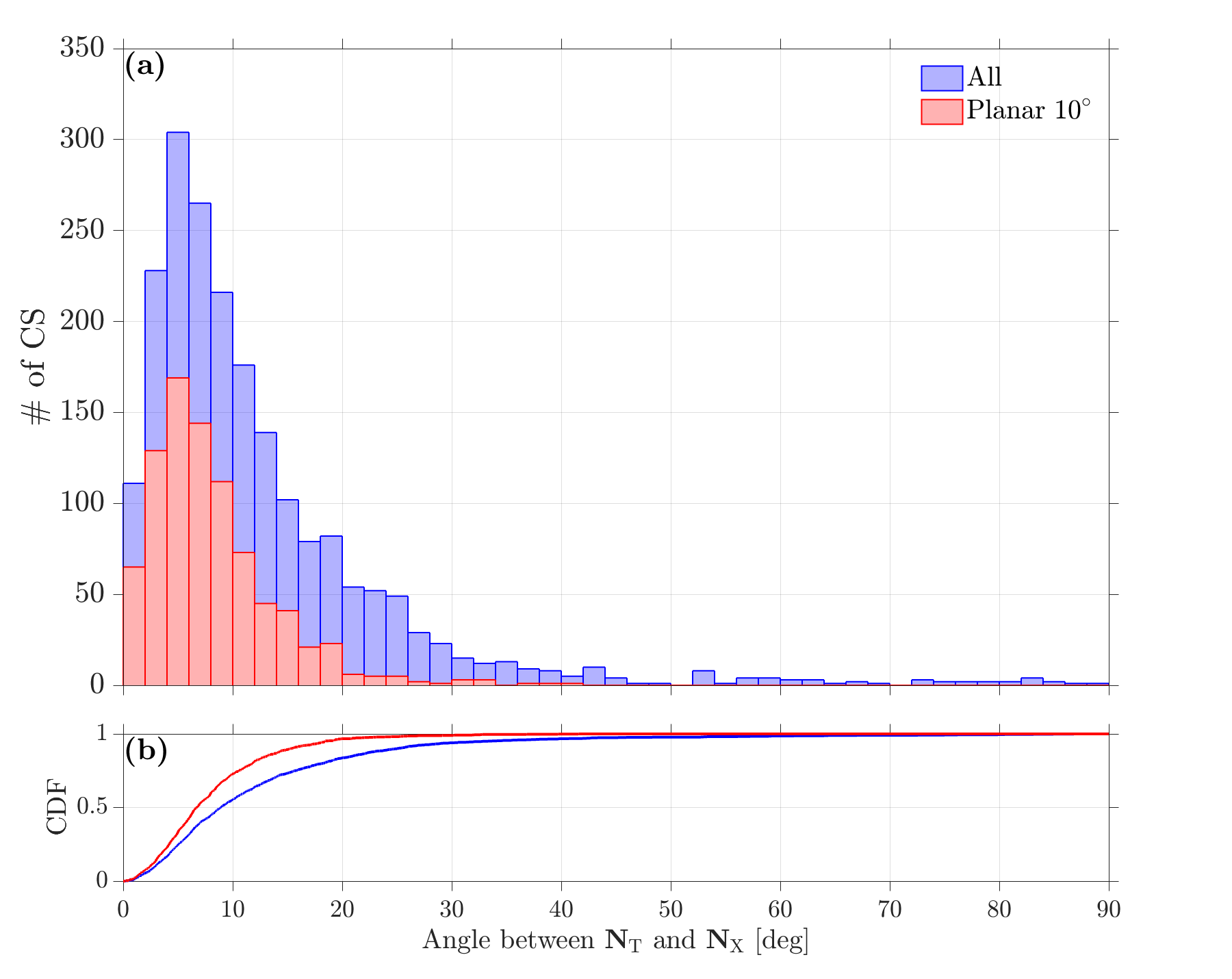}
\caption{The probability and cumulative distributions of the angle between timing normal ${\bf N}_{\rm T}$ and the cross-product vector ${\bf N}_{\rm X}$ that is a unit vector along the cross-product of magnetic fields at the CS boundaries. The distributions of different colors correspond to all CSs and CSs of sufficiently high planarity with $\alpha_{\rm max}\leq 10^{\circ}$.}
\label{fig5}
\end{figure}

\begin{figure}
\noindent\includegraphics[width=\textwidth]{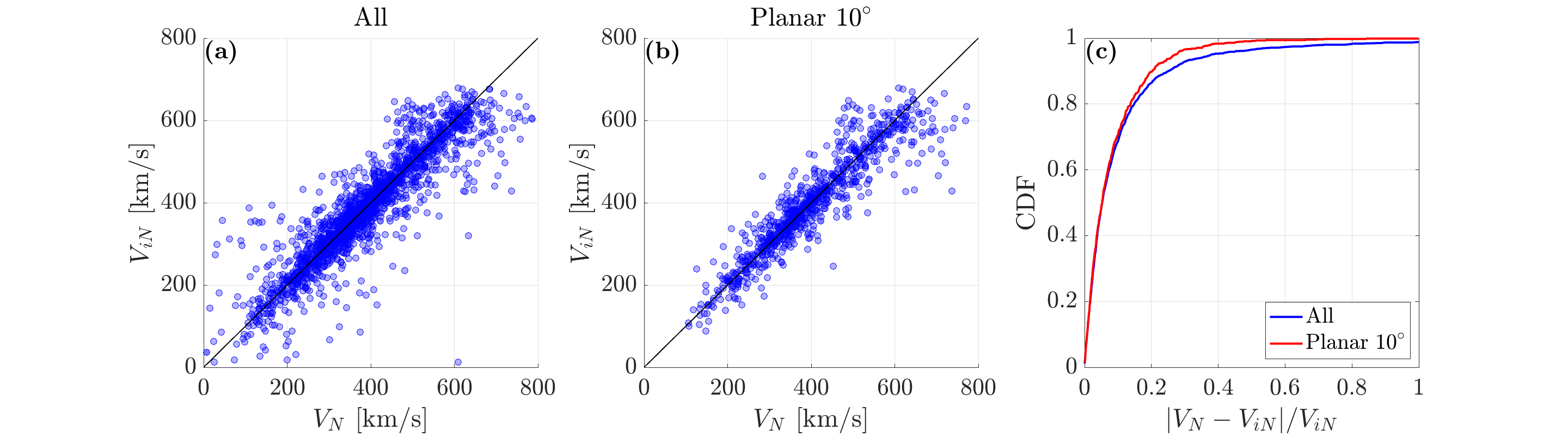}
\caption{The comparison between the CS propagation velocity $V_{N}$ revealed by the timing method and local ion flow velocity $V_{iN}$ along the timing normal. Panels (a) and (b) present scatter plots of $V_{iN}$ vs. $V_{N}$ for all CSs and CSs of sufficiently high planarity with $\alpha_{\rm max}\leq 10^{\circ}$, while panel (c) demonstrates corresponding cumulative distributions of $|V_{N}-V_{iN}|/V_{iN}$.}
\label{fig6}
\end{figure}

\begin{figure}
\noindent\includegraphics[width=\textwidth]{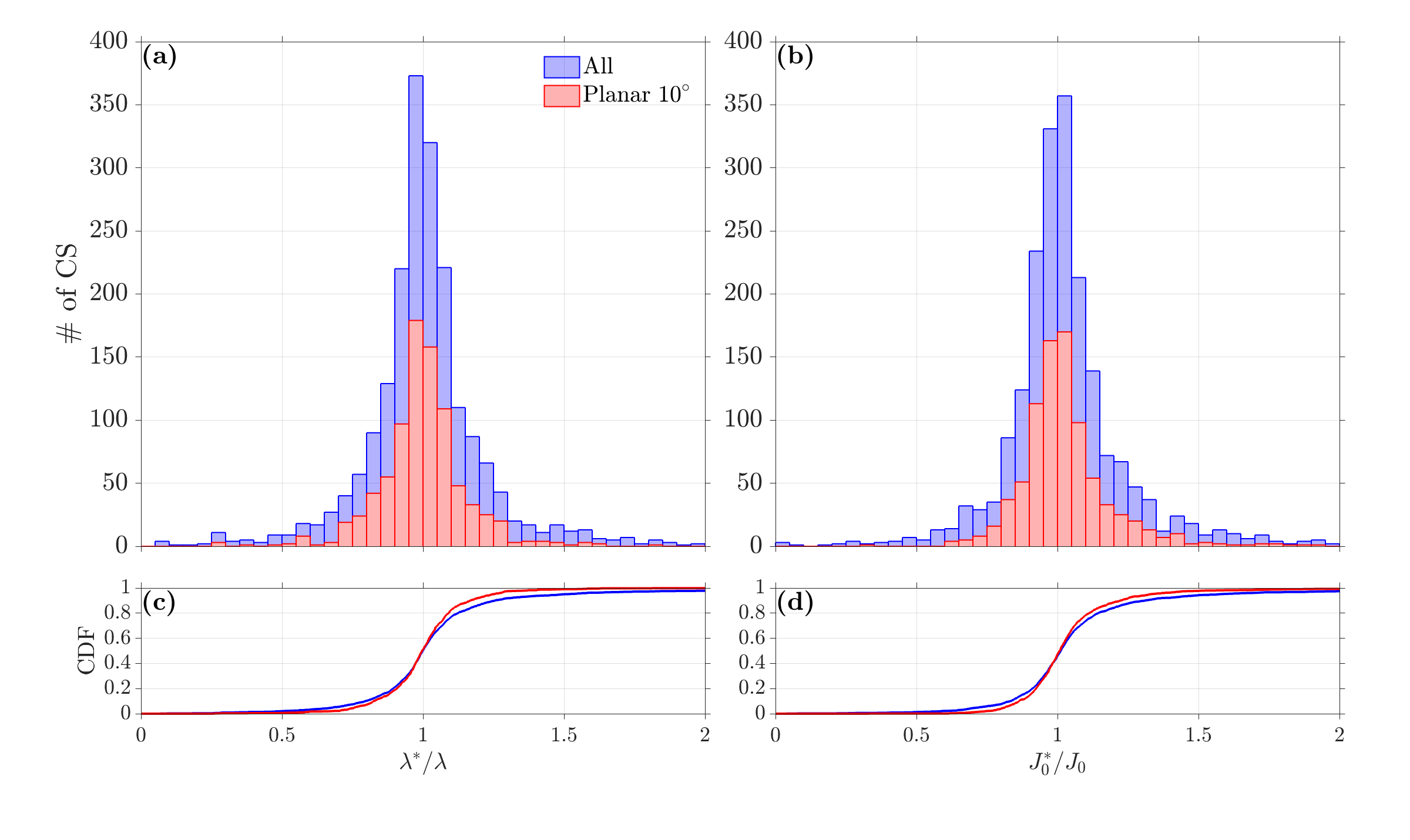}
\caption{The comparison between single-spacecraft and multi-spacecraft estimates of the CS thickness and current density. Single-spacecraft estimates $\lambda^{*}$ and $J_0^{*}$ were obtained by computing the CS normal using the cross-product method and assuming that the CSs are frozen into local plasma flow (Taylor hypothesis). Multi-spacecraft estimates $\lambda$ and $J_0$ were obtained using the CS normal and propagation velocity estimated by the timing method. The distributions of different colors correspond to all CSs and CSs of sufficiently high planarity with $\alpha_{\rm max}\leq 10^{\circ}$}
\label{fig7}
\end{figure}


Figures \ref{fig5} demonstrates probability and cumulative distributions of the angle between the cross-product vector ${\bf N}_{\rm X}$ and the timing normal ${\bf N}_{\rm T}$. The angle is within 20$^{\circ}$ for about 80\% of the CSs and the agreement is even better for the CSs of higher planarity. This better agreement results from a lower uncertainty of the timing normal for the CSs of higher planarity. The higher the planarity the more accurately the timing normal aligns with the actual CS normal and the better is the alignment between the timing normal and the cross-product vector. The fact that the cross-product vector delivers a reasonable estimate of the actual CS normal implies that the normal magnetic field component across the CSs is typically small compared to the guide field. The distribution of the angle between ${\bf N}_{\rm X}$ and ${\bf N}_{\rm T}$ for the CSs with $\alpha_{\rm max}\leq 10^{\circ}$ demonstrates that the cross-product method is rather accurate in estimating the actual CS normal; its angular uncertainty is within 15$^{\circ}$ (10$^{\circ}$) at the confidence level of 90\% (75\%).


Figure \ref{fig6} presents the comparison between local ion flow velocity along the timing normal $V_{iN}$ and the CS propagation velocity $V_{N}$ revealed by the timing method. Panels (a) and (b) show that the two velocities are typically consistent and agree even better for the CSs of higher planarity. Panel (c) shows $V_{N}$ and $V_{iN}$ are consistent within 20\% for more than 90\% of the CSs with $\alpha_{\rm max}\leq 10^{\circ}$. The better agreement between $V_{N}$ and $V_{iN}$ is certainly due to a lower uncertainty of the timing normal for the CSs of higher planarity. Thus, the Taylor hypothesis is statistically valid for CSs in the solar wind.

Figure \ref{fig7} presents the comparison of the half-thickness $\lambda$ and current density amplitude $J_0$ obtained using the four-spacecraft methodology (Figures \ref{fig2}) with corresponding estimates $\lambda^{*}$ and $J_0^{*}$ revealed by the single-spacecraft methodology based on the cross-product normal and frozen-in assumption. The four-spacecraft estimates have uncertainties associated with the degree of CS planarity on the scale of Cluster spacecraft separation, but these uncertainties are reduced by considering the CSs of sufficiently high planarity. Panels (a) and (b) demonstrate the ratio between four- and single-spacecraft estimates for the CSs with different $\alpha_{\rm max}$. The probability and cumulative distributions for the CSs with $\alpha_{\rm max}\leq 10^{\circ}$ demonstrate that the single-spacecraft methodology is quite accurate in estimating the CS thickness and current density amplitude. For more than about 90\% of the CSs with $\alpha_{\rm max}\leq 10^{\circ}$ we have $0.8\lesssim \lambda^{*}/\lambda\lesssim 1.2$ and $0.8\lesssim J_0^{*}/J_0\lesssim 1.2$. Thus, the single-spacecraft methodology based on the cross-product normal and frozen-in assumption delivers CS thickness and current density estimates with accuracy of better than 20\% at the confidence level of 90\%.

\section{Nature and properties of the current sheets\label{sec:4}}

Since solar wind CSs have negligible variations of the magnetic field magnitude \cite{Vasquez07,Vasko21:apj_rec}, whether a CS is a tangential or rotational discontinuity is determined solely by the magnitude of the normal magnetic field component. Note that the W\'{a}len relation well satisfied across solar wind CSs \cite{Paschmann13,Artemyev19:grl,Artemyev19:jgr} is not a unique feature of rotational discontinuities, since tangential discontinuities are not prohibited to satisfy it too \cite{Neugebauer06}. The normal component $B_{N}$ may be however within methodology uncertainty because of the angular uncertainty $\alpha_{\rm max}$ of the timing normal. Indeed, the timing normal is not likely to be strictly aligned with the actual CS normal and a significant fraction of the magnetic field component along the timing normal can be due to the projection of the actual guide field onto that normal. Because of the angular uncertainty, a finite component will be always observed along the timing normal even for tangential discontinuities. The nature of the CSs can be addressed by comparing the observed normal component $B_{N}$ with $\langle B\rangle \sin\alpha_{\rm max}$, where here and below $B_{N}$ and $\langle B\rangle$ are averaged values of the normal component and magnetic field magnitude over the CS central region. Note that $\langle B \rangle\sin\alpha_{\rm max}$ is the maximum value of the magnetic field component along the timing normal that would be observed across perfect tangential discontinuities because of the angular uncertainty $\alpha_{\rm max}$ of the timing method (SM).

\begin{figure}
\noindent\includegraphics[width=\textwidth]{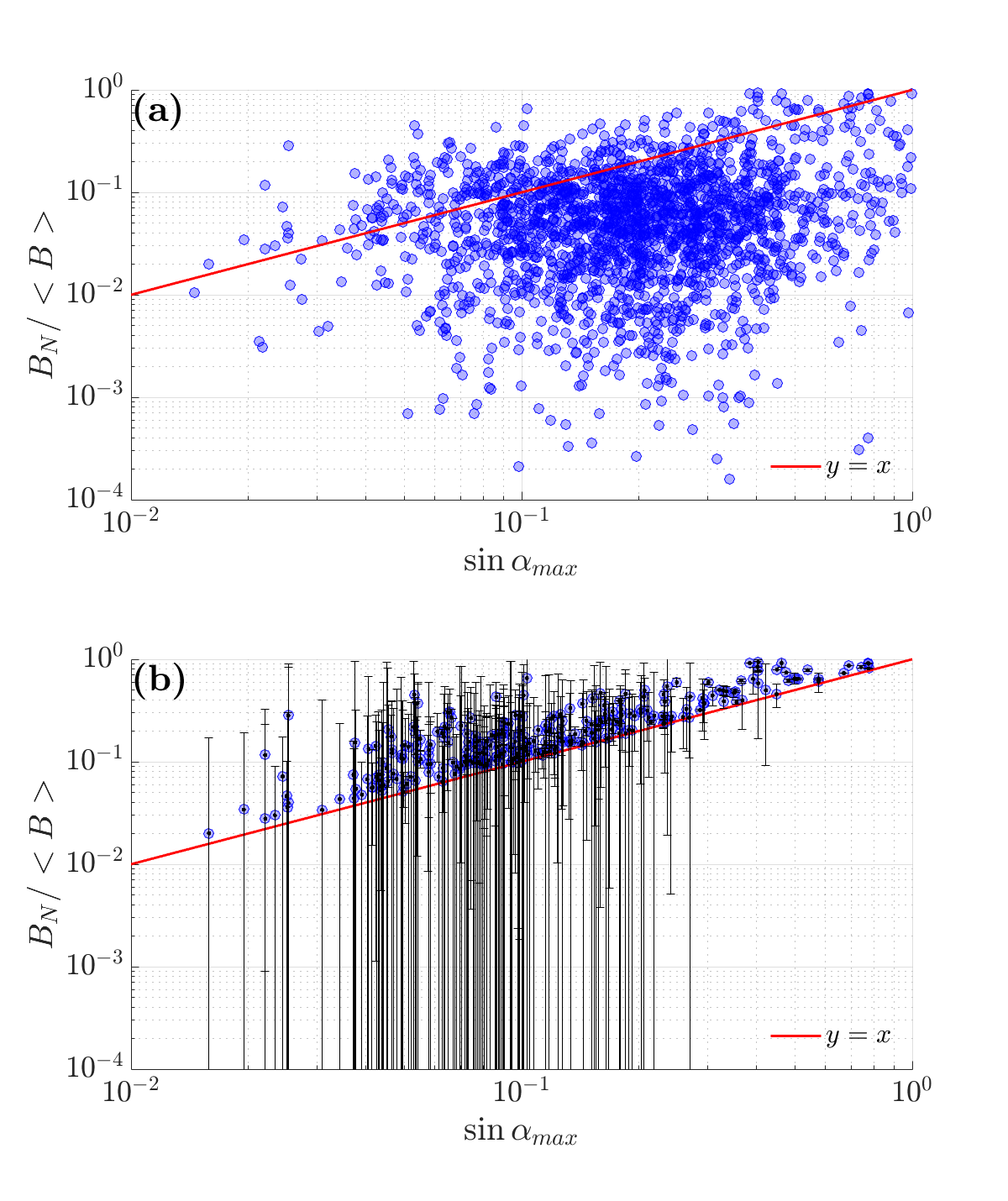}
\caption{The analysis of the magnetic field component along the timing normal: (a) a scatterplot of $B_{N}/\langle B\rangle$ versus its uncertainty $\sin\alpha_{\rm max}$ determined by the degree of CS planarity on the scale of Cluster spacecraft separation, where both the normal component $B_{N}$ and magnetic field magnitude $\langle B\rangle$ were averaged over the CS central region; (b) $B_{N}/\langle B\rangle$ versus $\sin\alpha_{\rm max}$ for the CSs with $B_{N}\gtrsim \langle B\rangle \sin\alpha_{\rm max}$, where error bars indicate the uncertainty caused by the uncertainty of the time delays used in the timing method (Section \ref{sec:4}).}
\label{fig8}
\end{figure}

Figure \ref{fig8} presents the analysis of the normal magnetic field component for all the collected 2,033 CSs. Panel (a) compares $B_{N}/\langle B\rangle$ and $\sin\alpha_{\rm max}$ and demonstrates that about 96\% of the CSs have the normal component within the uncertainty, $B_{N}\lesssim \langle B\rangle\sin \alpha_{\rm max}$. Thus, these CSs can be either tangential discontinuities or rotational discontinuities, whose normal component is within the methodology uncertainty. About 4\% of the CSs with $B_{N}\gtrsim \langle B\rangle\sin \alpha_{\rm max}$ could be interpreted in terms of rotational discontinuities if the angular uncertainty was the only uncertainty of the timing method. There is however an additional uncertainty caused by uncertainties of the time delays between the occurrence of the same CS aboard different Cluster spacecraft (Figure \ref{fig2}a). In the timing method we used time delays delivering the maximum cross-correlation between magnetic field profiles observed aboard different Cluster spacecraft, but each time delay has an uncertainty determined by the broadness of the cross-correlation function. We estimated uncertainties of the normal component caused by the uncertainty of the time delays for the CSs with $B_{N}\gtrsim \langle B\rangle \sin\alpha_{\rm max}$ (SM). Panel (b) presents the estimated uncertainties and demonstrates that a substantial fraction of this 4\% of the CSs may actually have $B_{N}\lesssim \langle B\rangle\sin\alpha_{\rm max}$, because of the relatively large uncertainty caused by the time delay uncertainties. Only half of those CSs or, equivalently, less than 2\% of the entire dataset have $B_{N}\gtrsim \langle B\rangle\sin\alpha_{\rm max}$. Thus, even four-spacecraft measurements do not allow us to classify the majority of the CSs in terms of tangential or rotational discontinuities. 

\begin{figure}
\noindent\includegraphics[width=\textwidth]{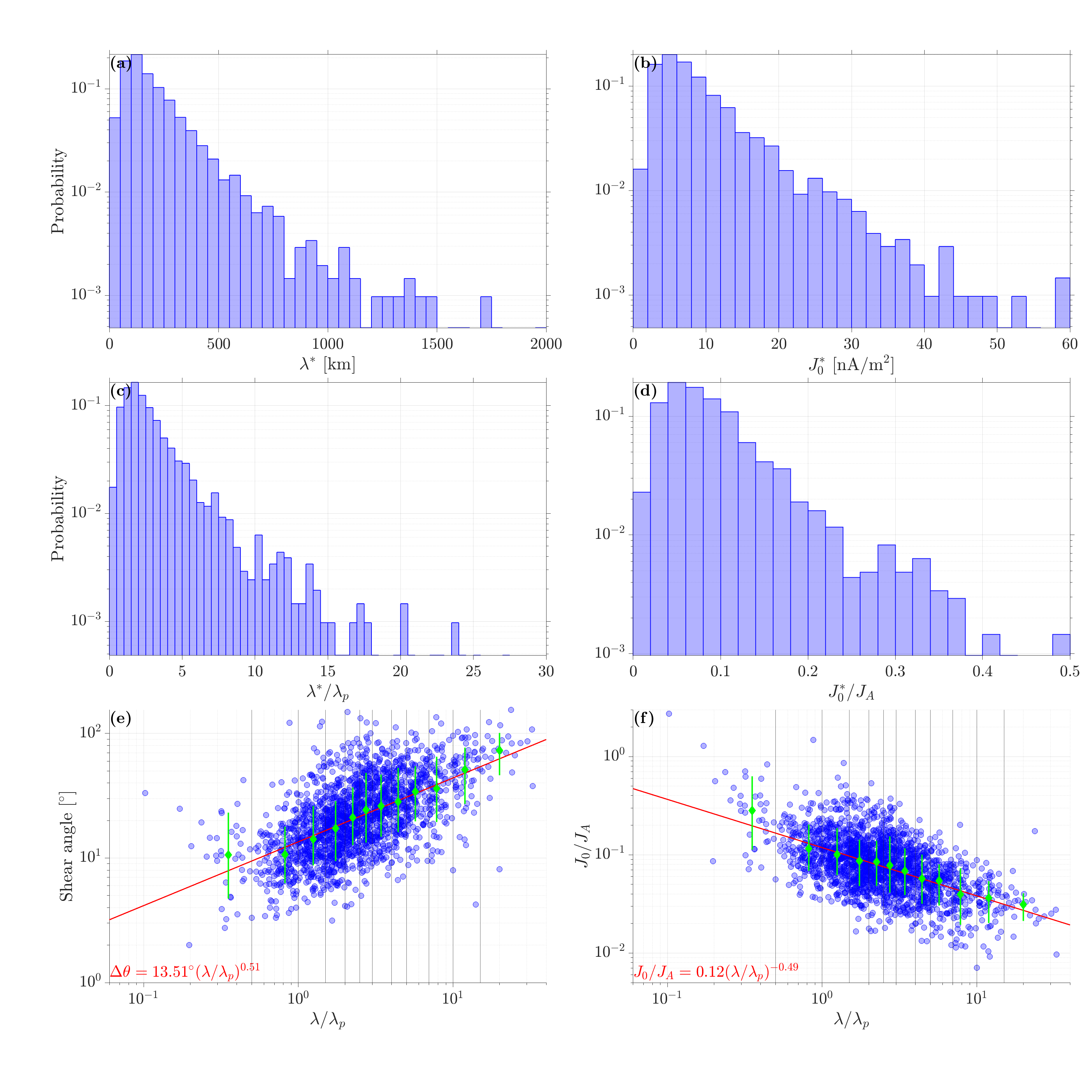}
\caption{The properties of the collected 2,033 CSs, except 28 bifurcated ones, estimated using the single-spacecraft methodology. Panels (a) and (b) show probability distributions of the CS half-thickness $\lambda^{*}$ and current density amplitude $J_0^{*}$ in physical units, while panels (c) and (d) present distributions of $\lambda^{*}/\lambda_{p}$ and $J_0^{*}/J_{A}$, where $\lambda_{p}$ is local proton inertial length and $J_{A}=en_{i}V_{A}$ is local Alfv\'{e}n current density. Panels (e) and (f) demonstrate the scale-dependence of the shear angle $\Delta \theta$ and normalized current density $J_0^{*}/J_{A}$ on normalized CS half-thickness $\lambda^*/\lambda_{p}$. The best power-law fits (red) and bin-mean profiles (green) are indicated in the panels. The bars reflect the range where 70\% of the data reside.}
\label{fig9}
\end{figure}

Figure \ref{fig9} presents the properties of all the collected CSs, except 28 bifurcated ones. We present single-spacecraft estimates $\lambda^{*}$ and $J_0^{*}$, since in contrast to multi-spacecraft estimates $\lambda$ and $J_0$ their uncertainty is independent of CS planarity as well as time delay uncertainties. Panels (a) and (b) demonstrate that the CSs have half-thickness from a few tens to a few thousand kilometers and current density amplitude from about a few to a few tens of nA/m$^{2}$. The most typical values of the CS half-thickness and current density are around 100 km and 5 nA/m$^{2}$, respectively. Panels (c) and (d) present the CS half-thickness and current density in local Alfv\'{e}n units and demonstrate that the CSs are proton kinetic-scale structures with current density amplitudes within a few tenths of local Alfv\'{e}n current density. The most typical values of the CS half-thickness and current density amplitude are around $1.5\lambda_{p}$ and $0.05J_{A}$, respectively. Panels (e) and (f) demonstrate that the shear angle $\Delta \theta$ and normalized current density $J_0^{*}/J_{A}$ are correlated with normalized half-thickness $\lambda^{*}/\lambda_{p}$. These scale-dependencies are revealed by bin median profiles as well as by power-law fits shown in the panels. The shear angle tends to be larger for larger-scale CSs, $\Delta \theta\approx 13.6^{\circ} (\lambda^{*}/\lambda_{p})^{0.5}$, while the current density amplitude tends to be larger for smaller-scale CSs, $J_0^{*}/J_{A}\approx 0.12\cdot (\lambda^{*}/\lambda_{p})^{-0.5}$. Note that the scale-dependence of the current density could be deduced from the scale-dependence of the shear angle, since $J_{0}^{*}\approx \langle B\rangle \Delta \theta/2\mu_0\lambda$ or, equivalently, $J_0^{*}/J_{A}\approx \Delta \theta \cdot (2\lambda/\lambda_{p})^{-1}$. These scale-dependencies and CS properties are discussed in the next section.

\section{Discussion}








The majority of studies of solar wind current sheets were performed using single-spacecraft measurements and relied on the normal estimates by MVA \cite{burlaga77,lepping86,Tsurutani96:jgr,soding01,Liu&Fu22}. Four-spacecraft analysis of solar wind CSs by Knetter et al., 2004 \cite{Knetter04} demonstrated however that MVA is not accurate in estimating the CS normal and the most accurate single-spacecraft estimate is delivered by the cross-product of magnetic fields at the CS boundaries. For that reason the recent statistical studies of magnetic reconnection \cite{Phan10,Phan20:apjs,Mistry17,Eriksson22:apj} and CSs in general \cite{Vasquez07,Vasko21:apj_rec,Vasko22:apj_origin,Lotekar22:apj} observed at 1 AU as well as near the Sun have been based on the cross-product normals. In this paper we have substantially expanded the previous multi-spacecraft studies of solar wind CSs using a dataset of more than two thousand CSs observed aboard four Cluster spacecraft. Our dataset is a factor of ten larger than the dataset of about a hundred CSs considered by Knetter et al., 2004 \cite{Knetter04} and includes CSs from 35 days, when the Cluster spacecraft were separated by either about 300 or 3,000 km (Table \ref{tab:list}). Four-spacecraft observations allowed us to estimate the CS normal and other CS properties using the timing method, while their comparison to different single-spacecraft estimates allowed revealing the accuracy of the latter. Note that the timing method has uncertainty determined by the CS planarity on the scale of Cluster spacecraft separation, but considering CSs of sufficiently high planarity allowed us to reduce this uncertainty. The use of the extensive dataset allows achieving statistical validity of all our conclusions.  



We \blue{showed} that MVA is indeed {\it highly inaccurate} in estimating the normal to CSs in the solar wind (Figure \ref{fig4}). Surprisingly, the minimum variance direction that is supposed to deliver the normal is often highly-oblique to the actual CS normal and, hence, more aligned with the actual guide field direction. Even more surprising is that the MVA accuracy is not correlated with the ratio $\lambda_{\rm int}/\lambda_{\rm min}$ between intermediate and minimum variances that is supposed to be a theoretical indicator of MVA accuracy. Importantly, the fundamental assumption of the MVA method is that the noise (instrumental or natural) is isotropic \cite{Behannon80:jgr,Sonnerup&Scheible98}. Only in that case the CS normal and guide field directions are delivered by respectively minimum and intermediate variance directions, while the ratio $\lambda_{\rm int}/\lambda_{\rm min}$ indicates the MVA accuracy. However, magnetic field fluctuations within and around CSs (the natural noise from the MVA perspective) are not isotropic, since solar wind turbulence is generally anisotropic. The magnetic field fluctuations in solar wind turbulence tend to be perpendicular to local magnetic field, $\delta B_{\perp}^2\gg \delta B_{\parallel}^2$ \cite{Smith06:jgr_anis,Hamilton08:jgr_anis,Chen16:jpp}. For that reason a CS with a small normal component (say, a perfect tangential discontinuity) immersed into solar wind turbulence may have the minimum variance along background magnetic field that is mostly along the guide field direction, rather than the actual CS normal. Thus, because of the natural anisotropy of solar wind turbulence the minimum variance direction delivered by MVA can be highly oblique to the actual CS normal in accordance with observations (Figure \ref{fig4}). In turn, MVA accuracy is not correlated with $\lambda_{\rm int}/\lambda_{\rm min}$ (Figure \ref{fig4}), because variances $\lambda_{\rm int}$ and $\lambda_{\rm min}$ are dominated by ambient anisotropic magnetic field fluctuations. Note that we also tried to apply MVA only to the magnetic field at the CS boundaries, rather than to the entire CS transition region, but we observed no improvement of the MVA accuracy (not shown). 



Since MVA is highly inaccurate in the solar wind, we agree with Knetter et al., 2004 \cite{Knetter04} and Neugebauer et al., 2006 \cite{Neugebauer06} that the previous classifications of solar wind CSs in terms of tangential or rotational discontinuities based on the magnetic field component along the MVA normal are not reliable \cite{burlaga77,lepping86,Tsurutani96:jgr,soding01}. The same statement is also relevant to the recent classification of solar wind current sheets \cite{Liu&Fu22} and current layers at the boundaries of switchbacks observed in the near-Sun solar wind \cite{Krasnoselskikh20:apj,Larosa21:aa,Akhavan21:aa}. In fact, even four-spacecraft observations do not allow us to confidently classify the current sheets in terms of tangential or rotational discontinuities. Note that the W\'{a}len relation well satisfied across solar wind CSs \cite{Paschmann13,Artemyev19:grl,Artemyev19:jgr} is not a unique feature of rotational discontinuities, since tangential discontinuities are not prohibited to satisfy it too \cite{Neugebauer06}. Thus, the classification of solar wind currents sheets in terms of tangential and rotational discontinuities remains a challenge.




We \blue{demonstrated} that the cross-product of magnetic fields at the CS boundaries delivers an accurate estimate of the actual CS normal that is in accordance with the analysis by Knetter et al., 2004 \cite{Knetter04}. The angular uncertainty of the cross-product method is less than 15$^{\circ}$ (10$^{\circ}$) at the confidence level of at least 90\% (75\%) (Figure \ref{fig5}). This high precision of the cross-product method implies that the normal component is typically much smaller than the guide field component. We also demonstrated that the Taylor frozen-in hypothesis is statistically valid for solar wind CSs that is in accordance with previous analysis of solar wind turbulence (e.g., Chasapis et al., 2017 \cite{Chasapis17:taylor}). The CS propagation velocity is consistent with local ion flow velocity within about 20\% at the 90\% confidence level (Figure \ref{fig6}). The single-spacecraft methodology based on the cross-product normal and frozen-in assumption has been widely used for the analysis of current sheets \cite{Vasquez07,Vasko21:apj_rec,Vasko22:apj_origin,Lotekar22:apj} and magnetic reconnection \cite{Phan10,Phan20:apjs,Gosling&Phan13,Mistry17,Eriksson22:apj} in the solar wind. We showed that this single-spacecraft methodology is actually accurate in estimating the CS thickness and current density amplitude; it delivers the CS properties within 20\% of their actual values at the confidence level of 90\%  (Figure \ref{fig7}). This result should be valuable for the analysis of current sheets near the Sun and beyond 1 AU, where only single-spacecraft measurements are available. The only unavoidable feature of this methodology is that it does not allow classifying CSs in terms of tangential or rotational discontinuities, because the normal component along the cross-product normal is by definition zero.

The collected CSs are proton kinetic-scale structures with half-thickness from a fraction to about ten proton inertial lengths and current density amplitude typically around 5\% of local Alfv\'{e}n current density. The most typical values of the half-thickness and current density for the CSs at 1 AU are around 100 km and 5 nA/m$^{2}$, which is consistent with previous reports \cite{Podesta17:jgr,Vasko21:apj_rec,Vasko22:apj_origin}. We also revealed that the shear angle and current density amplitude scale with thickness, $\Delta \theta\approx 13.6^{\circ} (\lambda^{*}/\lambda_{p})^{0.5}$ and $J_0^{*}/J_{A}\approx 0.12\cdot (\lambda^{*}/\lambda_{p})^{-0.5}$. Similar properties and almost identical scale-dependencies were reported for a larger dataset of CSs observed at 1 AU aboard Wind spacecraft \cite{Vasko21:apj_rec,Vasko22:apj_origin}. The scale-dependencies were shown to be a strong evidence supporting the hypothesis that the CSs are produced by turbulence cascade \cite{Vasko22:apj_origin}. The other evidence supporting this hypothesis is the similarity between distributions of waiting times of CSs observed in the solar wind and numerical turbulence simulations \cite{Vasquez07,Greco08,Greco09:apjl}. The revealed CS properties in Figure \ref{fig9} are basically consistent with those observed in turbulence simulations \cite{Franci17:apj,Papini19:apj_rec,Jain21:apj,Azizabadi21:phpl}, though a more detailed comparative analysis is still necessary. The fact that the normal to solar wind CSs is typically aligned with the cross-product of magnetic fields at the CS boundaries is also consistent with theory and numerical simulations that turbulence cascade produces CSs with wave vectors oriented perpendicular to background magnetic field \cite{Shebalin83:jpp,Boldyrev06:prl,Zhdankin13:apjl,Sisti21:aa}.


Even though there are several CSs in our dataset with half-thickness of 0.1$\lambda_{p}$, typically the CS half-thickness is above 0.3$\lambda_{p}$ (Figure \ref{fig9}). Note that electron-scale CSs could be resolved by Cluster magnetic field measurements, since the temporal resolution of 25 S/s translates into the spatial resolution of the order of 10 km that is around 0.1$\lambda_{p}$. The low occurrence of CSs thinner than about proton inertial length may indicate that CSs, whose thickness is smaller than this value, are efficiently disrupted by magnetic reconnection. This would be consistent with some turbulence and reconnection theories \cite{Loureiro&Boldyrev17,mallet17:mnras}. The alternative is that electron-scale CSs are abundant in the solar wind, but did not fit our criterion of relatively stable magnetic fields at the current sheet boundaries.

\section{Conclusion}

The results of this study can be summarized as follows

\begin{enumerate}
    \item The Minimum Variance Analysis is {\it highly inaccurate} in estimating the normal of current sheets in the solar wind. The minimum variance direction that is supposed to deliver the normal is often highly-oblique to the actual CS normal (the angle is larger than 60$^{\circ}$ for about 50\% of the CSs) and, hence, more aligned with the actual guide field direction. We believe that the MVA inaccuracy is caused by ambient turbulent fluctuations, which are typically anisotropic. \\
    
    \item The cross-product of magnetic fields at the current sheet boundaries delivers the current sheet normal with angular uncertainty of less than 15$^{\circ}$ (10$^{\circ}$) at the confidence level of at least 90\% (75\%). The current sheets are essentially frozen into plasma flows, since their propagation velocity is consistent with local ion flow velocity within about 20\% at the confidence level of 90\%.\\

    \item The single-spacecraft methodology based on the cross-product normal and frozen-in assumption delivers the current sheet thickness and current density amplitude within 20\% of their actual values at the confidence level of 90\%. This methodology will be valuable for the analysis of current sheets near the Sun and beyond 1 AU, where only single-spacecraft measurements are available.\\

    \item Solar wind current sheets at 1 AU are proton kinetic-scale structures with the most typical half-thickness around 100 km that is around proton inertial length and the most typical current density amplitude of 5 nA/m$^2$ that is around 5\% of the local Alfv\'{e}n current density. Solar wind current sheets have the shear angle and current density amplitude scaling with their thickness, $\Delta \theta\approx 13.5^{\circ} (\lambda/\lambda_{p})^{0.5}$ and $J_0/J_{A}=0.12 (\lambda/\lambda_{p})^{-0.5}$, where $\lambda_{p}$ and $J_{A}=en_{i}V_{A}$ are local proton inertial length and Alfv\'{e}n current density.\\
    
    \item The normal magnetic field component is within the methodology uncertainties and classifying solar wind current sheets in terms of tangential and rotational discontinuities remains a challenge even using multi-spacecraft measurements.
    
\end{enumerate}

{\bf Acknowledgments:}
We would like to acknowledge Cluster Active Archive and Cluster instrument teams, in particular FGM, CIS/CODIF and CIS/HIA, Whisper for excellent data.  The work of T.P. was supported by NASA Living With a Start grant No. 80NSSC20K1781. The work of F.M. was supported by NASA Heliophysics Guest Investigator grant No. 80NSSC21K0730. I.V. thanks Russian Science Foundation grant No. 21-12-00416. 

\newpage
 \begin{table}
 \caption{The summary of the dataset of 2,033 current sheets collected in the solar wind. For each day indicated in the first column we analyzed several time intervals (SM). The next four columns present the magnetic field magnitude $B$, plasma density $n_{i}$, plasma flow velocity $V_{i}$, and proton beta $\beta_{p}$ all averaged over the considered time intervals. The last three columns present the maximum and minimum spatial separation ($\Delta R_{\rm max}$ and $\Delta R_{\rm min}$) between Cluster spacecraft and the number of CSs collected within the time intervals for each day. }
 \centering
 \begin{tabular}{|c|c|c|c|c|c|c|c|}
 \hline
  Date  & $B$ [nT] & $n_{i}\;[{\rm cm}$$^{-3}$$]$ & $V_{i}$ [km/s] & $\beta_{p}$ & $\Delta R_{\rm max}$ [km] & $\Delta R_{\rm min}$ [km] & CS\#  \\
 \hline
   20030108  & 5.2  & 22.9 & 280 & 0.07 & 4815 & 3210 & 17   \\
    \hline
   20030113  & 8.2  & 6.8  & 390 & 0.09 & 6345 & 2875 & 29   \\
    \hline
   20030122  & 8.3  & 6.2  & 600 & 0.13 & 4205 & 3385 & 96   \\
    \hline
   20030123  & 8.8  & 5.1  & 645 & 0.17 & 6040 & 2920 & 33   \\   
    \hline
   20030129  & 9.6  & 25.5 & 430 & 0.19 & 4110 & 3460 & 5    \\
    \hline
   20030130  & 11.7 & 9.7  & 455 & 0.47 & 6780 & 3015 & 118  \\
    \hline
   20030210  & 7.0  & 7.4  & 440 & 0.40 & 4230 & 3350 & 42   \\
    \hline
   20030211  & 6.9  & 9.0  & 420 & 0.39 & 8100 & 3090 & 46   \\
    \hline
   20030212  & 5.6  & 16.1 & 340 & 0.31 & 4800 & 3350 & 37   \\
    \hline
   20030213  & 10.7 & 11.9 & 365 & 1.2  & 5000 & 3240 & 119  \\
    \hline
   20030217  & 8.7  & 5.0  & 620 & 0.87 & 4000 & 3460 & 68   \\
    \hline
   20030218  & 13.4 & 8.2  & 660 & 2.4  & 4595 & 3220 & 72   \\
    \hline
   20030219  & 6.4  & 7.2  & 545 & 1.3  & 4910 & 3295 & 10   \\
    \hline
   20030220  & 8.3  & 7.6  & 585 & 1.7  & 4695 & 3285 & 63   \\
    \hline
   20030225  & 6.3  & 8.4  & 405 & 0.25 & 8245 & 2785 & 27   \\
    \hline
   20030226  & 14.3 & 30.5 & 440 & 0.25 & 5240 & 3785 & 5    \\
    \hline
   20030227  & 10.1 & 7.3  & 550 & 1.4  & 4330 & 3375 & 75   \\
    \hline
   20030301  & 7.1  & 6.6  & 415 & 0.14 & 4255 & 3435 & 29   \\
    \hline
   20030302  & 5.6  & 8.8  & 415 & 0.41 & 7630 & 2980 & 9    \\
    \hline
   20030311  & 7.5  & 10.7 & 410 & 0.57 & 4390 & 3300 & 104  \\
    \hline
   20030313  & 11.1 & 9.9  & 485 & 0.21 & 4015 & 3355 & 20   \\ 
    \hline
   20030314  & 10.9 & 12.3 & 535 & 0.79 & 8110 & 2765 & 10   \\ 
    \hline
   20030325  & 4.6  & 7.5  & 415 & 0.49 & 4495 & 3285 & 97   \\
    \hline
   20030326  & 6.1  & 15.7 & 385 & 0.95 & 8065 & 2795 & 17   \\
    \hline
   20030329  & 9.0  & 21.6 & 430 & 0.02 & 5105 & 3455 & 13   \\
    \hline
   20030330  & 10.4 & 16.9 & 450 & 0.29 & 4525 & 3295 & 66   \\
    \hline
   20031219  & 2.7  & 13.0 & 320 & 0.52 & 220  & 190  & 9   \\
    \hline
   20031220  & 8.7  & 37.5 & 340 & 4.9  & 325  & 165  & 71   \\
    \hline
   20031231  & 8.4  & 18.4 & 440 & 2.8  & 240  & 155  & 44   \\
    \hline
   20040102  & 8.3  & 12.9 & 460 & 0.82 & 240  & 160  & 159  \\
    \hline
   20040122  & 18.3 & 23.6 & 645 & 9.1  & 235  & 180  & 188  \\
    \hline
   20040124  & 8.8  & 9.8  & 505 & 0.76 & 230  & 175  & 83   \\
   \hline
   20040126  & 11.0 & 10.3 & 445 & 0.63 & 215  & 190  & 62   \\
    \hline
   20040127  & 9.9  & 7.2  & 430 & 0.22 & 235  & 175  & 23   \\
    \hline
   20040131  & 8.5  & 4.8  & 570 & 0.44 & 240  & 170  & 174  \\
 \hline
 \end{tabular}
 \label{tab:list}
 \end{table}

\bibliographystyle{unsrt}

\end{document}